\documentclass[12pt]{iopart}

\usepackage{cite}
\usepackage{iopams}
\usepackage{epsfig}

\begin{document}

\title[The quantum $J_{1}$--$J_{1}'$--$J_{2}$ spin-$1/2$ Heisenberg model]{The quantum $J_{1}$--$J_{1}'$--$J_{2}$ spin-$1/2$ Heisenberg model: 
Influence of the interchain coupling on the ground-state magnetic ordering in 2D}

\author{R F Bishop$^{1,2}$, P H Y Li$^{1,2}$, R Darradi$^{3}$ and J Richter$^{3}$}
\address{$^{1}$ School of Physics and Astronomy, Schuster Building, The University of Manchester, Manchester, M13 9PL, UK}
\address{$^{2}$ School of Physics and Astronomy, University of Minnesota, 116 Church Street SE, Minneapolis, Minnesota 55455, USA}
\address{$^{3}$ Institut f\"ur Theoretische Physik, Universit\"at Magdeburg, 39016 Magdeburg, Germany}

\begin{abstract}
We study the phase diagram of the 2D $J_{1}$--$J_{1}'$--$J_{2}$
spin-$1/2$ Heisenberg model by means of the coupled cluster method. The effect
of the coupling $J_{1}'$ on the N\'{e}el and stripe states is
investigated.  We find that the quantum critical points for the
N\'{e}el and stripe phases increase as the coupling strength 
$J_{1}'$ is increased, and an intermediate phase emerges above the region at
$J_{1}'\approx0.6$ when $J_{1}=1$.  We find indications for a quantum 
triple point at $J_{1}'\approx0.60\pm0.03$, $J_{2} \approx 0.33 \pm 0.02$ for $J_{1}=1$.
\end{abstract}


\section{Introduction}
The spin-$1/2$ Heisenberg antiferromagnet has been much
studied in recent years due to the discovery or successful
syntheses of such new magnetic materials as the layered-oxide
high-temperature superconductors.  Also of much interest has been the
interplay between frustration and quantum fluctuations in
$2$D quantum spin systems that can lead to quantum phase transitions
between magnetically ordered semiclassical and novel quantum
paramagnetic ground-state phases, see, e.g., Refs.~\cite{Ri:2004,Mis:2005}.

For example, the frustrated 2D antiferromagnetic $J_{1}$--$J_{2}$ model with nearest-neighbour ($J_{1}$) and
next-nearest-neighbour ($J_{2}$) bonds has attracted much attention both
theoretically (see, e.g., Refs.~\cite{Ch:1988,Ri:1993,Schulz:1996,Bi:1998,Ca:2000,Ro:2004,
Sir:2006,Schm:2006} and references cited therein) and 
experimentally~\cite{Mel:2000,Car:2002}.  It is now well accepted that the model exhibits two 
phases displaying
semiclassical magnetic long-range order (LRO) at small and at large $J_2$,
separated by an intermediate quantum paramagnetic phase without
magnetic LRO in the parameter region
$J_{c_1}<J_2<J_{c_2}$ where $J_{c_1} \approx 0.4J_{1}$ and $ J_{c_2}
\approx 0.6J_{1}$.  The ground state (GS) for $J_2<J_{c_1}$ exhibits
N\'{e}el magnetic LRO, whereas for $J_2>J_{c_2}$ it exhibits collinear stripe LRO.  

In real systems deviations from the ideal 2D $J_{1}$--$J_{2}$ model such as spin
anisotropies~\cite{Ro:2004,Via:2007} or interlayer coupling may be
relevant~\cite{Schm:2006}.  In addition, a 3D version of the $J_{1}$--$J_{2}$ model~\cite{Schmidt:2002} has also been considered.

An interesting generalization of the pure $J_1$--$J_2$ model has
been introduced recently by Nersesyan and Tsvelik~\cite{Ne:2003}.
They consider a spatially anisotropic spin-$1/2$ 2D
$J_{1}$--$J_{1}'$--$J_{2}$ model, where the nearest-neighbour bonds
have different strengths $J_{1}$ and $J_{1}'$ in the $x$ and $y$ directions.  This
model has been further studied by other groups using the exact diagonalization (ED) of small lattice samples of
$N\leq36$ sites~\cite{Si:2004}, and the continuum limit of the
model~\cite{Star:2004}.  Both groups support the prediction of the resonating valence
bond state by Nersesyan and Tsvelik~\cite{Ne:2003} for
$J_{2}=0.5J_{1}' \ll J_{1}$, and the limit of small interchain coupling
extends along a curve nearly coincident with the line where the energy
is maximum.  The model has also been studied by
Moukouri~\cite{Mo:2006} using a two-step density-matrix
renormalization group approach.
  
Our aim here is to further the study of this model by using the coupled cluster method
(CCM).  The CCM (see, e.g., Refs.~\cite{Bi:1991,Bi:1998_b,Fa:2004}
and references cited therein) is one of the most powerful and
universally applicable techniques of quantum many-body theory.  It has
been applied successfully to calculate with high accuracy the
ground- and excited- state properties of many lattice quantum spin
systems (see, e.g.,
Refs.~\cite{Schm:2006,Fa:2004,Ze:1998,Kr:2000,Bi:2000,Fa:2002,Dar:2005} and references
cited therein). 
A particularly important result from our
calculations is the indicated existence of a quantum triple point (QTP) at
nonzero (positive) values of $J_{1}$, $J_{1}'$ and $J_{2}$.

\section{The model}
\begin{figure}[b]
\begin{center}
\epsfig{file=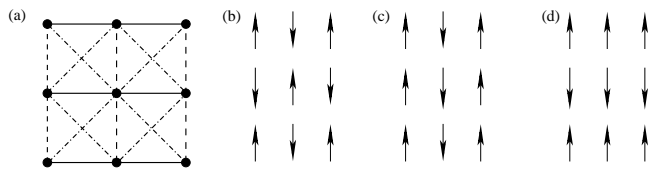,width=8.5cm}
\caption{(a) $J_{1}$--$J_{1}'$--$J_{2}$ model; --- $J_{1}$; - - $J_{1}'$; $\cdot - \cdot$ $J_{2}$; (b) N\'{e}el state, 
(c) stripe state - columnar
and (d) stripe state - row. Arrows in (b), (c) and (d) represent spins situated on the sites of the square lattice (indicated by $\bullet$ in (a)). }
\label{fig1}
\end{center}
\end{figure}

The $J_{1}$--$J_{1}'$--$J_{2}$ model is a spin-$1/2$ Heisenberg model
on a square lattice with three kinds of exchange bonds, with strength $J_{1}$ along the row direction, $J_{1}'$ along the column direction, and $J_{2}$ along
the diagonals, as shown in figure~\ref{fig1}.  All
exchanges are assumed positive here, and we set $J_{1} = 1$.
The Hamiltonian of the model is described by
\begin{eqnarray}
H &=& J_{1}\sum_{i,l}{\bf s}_{i,l}\cdot{\bf s}_{i+1,l} + J_{1}'\sum_{i,l}{\bf s}_{i,l}\cdot{\bf
s}_{i,l+1}\nonumber\\
 &+& J_{2}\sum_{i,l}({\bf s}_{i,l}\cdot{\bf s}_{i+1,l+1} + {\bf s}_{i+1,l} \cdot {\bf
s}_{i,l+1}),  \label{H}
\end{eqnarray}
where the index $(i,l)$ labels the $x$ (row) and $y$ (column) components of the lattice sites.

This model has two types of classical GS, namely, the
N\'{e}el ($\pi,\pi$) state and stripe states (columnar stripe
($\pi,0$) and row stripe ($0,\pi$)), the spin orientations of which
are shown in figures~\ref{fig1}(b,c,d) respectively.  There is clearly a symmetry
under the interchange of rows and columns, $J_{1} \rightleftharpoons
J_{1}'$, which implies that we need only consider the range of
parameters with $J_{1}'<J_{1}$.  The ground-state (gs) energies of the
various classical states are given by
\begin{eqnarray}
\frac{E^{\mbox{\tiny{cl}}}_{\mbox{\tiny{N\'{e}el}}}}{{N}} &=& \frac{1}{4}(-J_{1}-J_{1}'+2J_{2}),\nonumber\\
\frac{E^{\mbox{\tiny{cl}}}_{\mbox{\tiny{columnar}}}}{{N}} &=& \frac{1}{4}(-J_{1}+J_{1}'-2J_{2}),\nonumber \\ 
\frac{E^{\mbox{\tiny{cl}}}_{\mbox{\tiny{row}}}}{{N}} &=& \frac{1}{4}(J_{1}-J_{1}'-2J_{2}). \label{H_classical}
\end{eqnarray}
We take $J_{1}=1$ and $J_{1}'<1$.  Clearly, from (\ref{H_classical}), the classical GS is then either the
N\'{e}el state (if $J_{1}'>2J_{2}$) or the columnar stripe state (if
$J_{1}'<2J_{2}$).  Hence, the (first-order) classical phase transition between the
N\'{e}el and columnar stripe states occurs at
$J^{c}_{2}=\frac{1}{2}J_{1}',\,\forall J_{1}>J'_{1}$.

\section{The coupled cluster method}
\label{CCM}
The CCM formalism is now briefly described (and see
Refs.~\cite{Bi:1991,Bi:1998_b,Fa:2004,Ze:1998,Kr:2000,Bi:2000,Fa:2002,Dar:2005}
for further details).  The starting point for the CCM calculation is to select a normalized
model state $|\Phi\rangle$.  It is often convenient to take the classical ground state as the
model state for spin systems.  Hence our model states are the N\'{e}el state and the
columnar stripe state.   In order to treat each site identically, we perform a mathematical rotation of the local axes of the spins such that all spins in the reference state align along the negative $z$-axis.  
The Schr\"{o}dinger ground-state ket and bra CCM equations are $H|\Psi\rangle = E|\Psi\rangle$ and $\langle\tilde{\Psi}|H=E\langle\tilde{\Psi}|$ respectively.  The CCM employs the exponential ansatz, $|\Psi\rangle=$e$^{S}|\Phi\rangle$ 
and $\langle\tilde{\Psi}|=\langle\Phi|\tilde{\cal S}$e$^{-S}$. The correlation operator $S$ is expressed as $S = \sum_{I\neq0}{\cal
S}_{I}C^{+}_{I}$ and its counterpart is $\tilde{S} = 1
+ \sum_{I\neq0}\tilde{\cal S}_{I}C^{-}_{I}$.  The multispin creation operators $C^{+}_{I} \equiv (C^{-}_{I})^{\dagger}$, with $C^{+}_{0} \equiv 1$, are written as 
\(C^{+}_{I}\equiv s^{+}_{j_{1}} s^{+}_{j_{2}} \cdots s^{+}_{j_{n}}\).
The gs energy is $E = \langle\Phi|\mbox{e}^{-S}H\mbox{e}^{S}|\Phi\rangle$; and the staggered 
magnetization $M$ in the rotated spin 
coordinates is $M \equiv -\frac{1}{N} \langle\tilde{\Psi}|\sum_{j=1}^{N}s^{z}_{j}|\Psi\rangle$.  

The ket- and bra-state correlation coefficients $({\cal S}_{I}, \tilde{{\cal S}_{I}})$ are calculated by requiring the expectation value $\bar{H} \equiv \langle\tilde{\Psi}|H|\Psi\rangle$ to be a minimum with respect to all parameters $({\cal S}_{I}, \tilde{{\cal S}_{I}})$ such that $\langle \Phi|C^{-}_{I}\mbox{e}^{-S}H\mbox{e}^{S}|\Phi\rangle = 0$ and $\langle\Phi|\tilde{S}(\mbox{e}^{-S}H\mbox{e}^{S} - E_{0})C^{+}_{I}|\Phi\rangle = 0\;; \forall I \neq 0$. 

\section{Approximation scheme}
The CCM formalism is exact if all spin configurations are included in
the $S$ and $\tilde{S}$ operators.  In practice, however, truncations are needed.  As in much of our previous work we employ here the localized 
LSUB$n$ scheme,~\cite{Bi:1998,Fa:2004,Ze:1998,Kr:2000,Bi:2000,Fa:2002,Dar:2005,Schm:2006} in which all possible multi-spin-flip correlations over different locales on the lattice defined by $n$ or fewer contiguous lattice sites are retained.  The numbers of such fundamental configurations (viz., those that are distinct under the symmetries of the Hamiltonian and of the model state $|\Phi\rangle$) that are retained for the N\'{e}el and stripe states of the current model in 
various LSUB$n$ approximations are shown in Table~\ref{table_FundConfig}.  In order to solve the corresponding coupled sets of CCM bra- and ket- equations we use parallel computing~\cite{ccm}.
\begin{table}
\caption{Number of fundamental LSUB$n$ configurations ({$\sharp$ f.c.}) for the N\'{e}el and stripe states of the spin-$1/2$ $J_{1}$--$J_{1}'$--$J_{2}$ model.}
\label{table_FundConfig}
\begin{center}
\item[]\begin{tabular}{ccc} 
\br
{Method} & \multicolumn{2}{c}{$\sharp$ f.c.} \\ 
& N\'{e}el & stripe \\ 
\mr
LSUB$2$ & 2 & 1 \\ 
LSUB$4$ & 13 & 9 \\ 
LSUB$6$ & 146 & 106 \\ 
LSUB$8$ & 2555 & 1922 \\ 
LSUB$10$ & 59124 & 45825 \\ 
\br
\end{tabular} 
\end{center}
\end{table}

\section{Extrapolation scheme}
In practice one needs to extrapolate the raw LSUB$n$ data to
the $n\rightarrow\infty$ limit.  Based on previous experience~\cite{Schm:2006,Kr:2000,Dar:2005} we 
use the following empirical scaling laws for the
extrapolations of the gs energy per spin,
\begin{equation}
E/N=a_{0}+a_{1}n^{-2}+a_{2}n^{-4}\;,  \label{Extrapo_E}
\end{equation}
and the staggered magnetization of frustrated models,
\begin{equation}
M=b_{0}+b_{1}n^{-\nu} \;, \label{Extrapo_1}
\end{equation}
where the exponent $\nu$ is a fitting parameter.  In order to fit to any fitting formula that contains $n$ unknown parameters, 
one desirable rule is to have at least ($n+1$) data points ($n+1$ rule)
to obtain a robust and
stable fit.  
In our results below the LSUB$n$ results for $n=\{4,6,8,10\}$ are extrapolated.

\section{Results}
\begin{figure}[!htbp]
\begin{center}
\vskip0.2cm
\epsfig{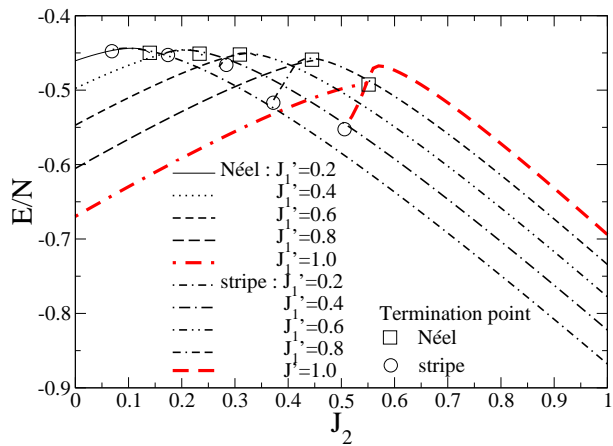}  
\caption{Extrapolated CCM LSUB$n$ results for the gs energy per spin, E/N, for $J_{1}'=0.2,0.4,0.6,0.8,1.0$, using the N\'{e}el and stripe states of the $s=1/2$ $J_{1}$--$J_{1}'$--$J_{2}$ model.  The LSUB$n$ results are extrapolated in the limit $n \rightarrow \infty$ using the set $n=\{4,6,8,10\}$.  The NN exchange coupling $J_{1}=1$.}
\label{fig2}
\end{center}
\end{figure}
In figure~\ref{fig2} we show the gs energy per spin as a function of
$J_{2}$.  For each value of $J_{1}'$ two curves are
shown, one (for smaller values of $J_{2}$) using the N\'{e}el state,
and the other (for larger values of $J_{2}$) using the stripe state as
CCM model state.  Both sets of curves have the natural termination
points~\cite{Bi:1998,Fa:2004,Ze:1998} shown.  For $J_{1}'$$\lesssim
0.6$ the two curves for a given value of $J_{1}'$ cross (or, in the
limit, meet) very smoothly near their maxima, at a value of $J_{2}$
slightly larger than the classical transition point of $0.5J_{1}'$.
This behaviour is indicative of a second-order quantum phase
transition, by contrast with the first-order classical transition from
 (\ref{H_classical}).  For $J_{1}'\gtrsim0.6$ the curves no longer
cross at a physical value (viz., where the calculated staggered
magnetization is positive), indicating the opening up of an
intermediate quantum phase between the semiclassical N\'{e}el and stripe phases.

\begin{figure}[!htbp]
\begin{center}
\epsfig{file=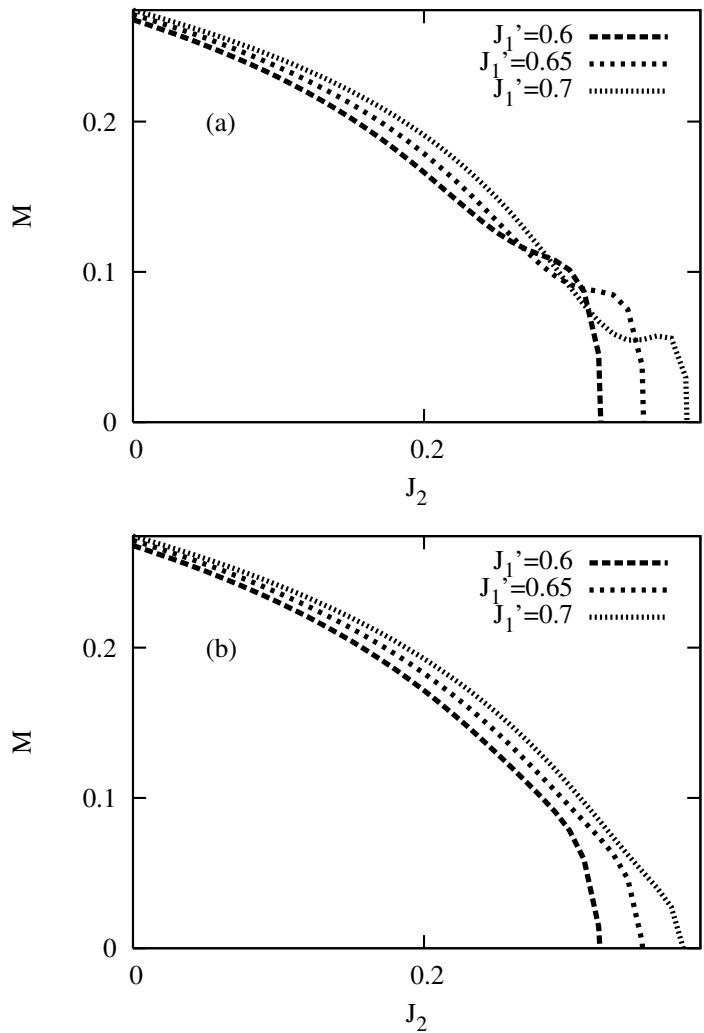,width=14cm,angle=270}   
\caption{Extrapolated CCM LSUB$n$ results for the gs staggered magnetization, $M$, for 
$J_{1}'=0.6,0.65,0.7$
for the N\'{e}el state of the $s=1/2$ $J_{1}$--$J_{1}'$--$J_{2}$ model.  
(a) Results using (\ref{Extrapo_1}), \(M=b_{0}+b_{1}n^{-\nu}\).
(b) Results using (\ref{Extrapo_4}), \(M=c_{0}+n^{-0.5}\left(c_{1}+c_{2}n^{-1}\right)\). The LSUB$n$ results are extrapolated in the limit $n \rightarrow \infty$ using the set $n=\{4,6,8,10\}$.  The NN exchange coupling $J_{1}=1$.}
\label{fig3}
\end{center}
\end{figure}

In figure~\ref{fig3}(a) 
we show the equivalent staggered magnetization,
$M$, for the N\'{e}el state, extrapolated using (\ref{Extrapo_1}).
We observe that the extrapolation scheme produces smooth and
physically reasonable results, except for a very narrow anomalous ``shoulder'' region near the points where $M$ vanishes for
$0.6\lesssim$$J_{1}'$$\lesssim0.75$.  This critical regime is
undoubtedly difficult to fit with the simple two-term scheme of
 (\ref{Extrapo_1}).  Our method for curing this problem and
stabilizing the curves is to make efficient use of the information we
obtain in (\ref{Extrapo_1}) to extract the exponent $\nu$, and
then to use that value to infer the next term in the series.  We find,
very gratifyingly, that the value for $\nu$ fitted to (\ref{Extrapo_1}) turns out to be very close to $0.5$ for
all values of $J_{1}'$ and $J_{2}$ except very close to the critical
point.  Therefore, we use the form of (\ref{Extrapo_4}),
\begin{equation}
M=c_{0}+n^{-0.5}\left(c_{1}+c_{2}n^{-1}\right) \;.
\label{Extrapo_4}
\end{equation}
The use of (\ref{Extrapo_4}) now removes the
anomalous shoulder, as shown in
figure~\ref{fig3}(b).  Henceforth, in all of the results we discuss, we
use (\ref{Extrapo_4}) for the staggered magnetization.

The raw LSUB$n$ data for $n$=\{2,4,6,8,10\} and $n$=\{6,8,10\} are also extrapolated.  The results for $n$=\{2,4,6,8,10\} and 
$n$=\{4,6,8,10\} are similar as they both obey the $n+1$ rule as mentioned above.  This adds credence to the validity and stability of our results.  
The extrapolated curve for $n=\{6,8,10\}$ has a very minor ``shoulder'' which is undoubtedly due to using only three data points to fit the three unknown terms, thus violating the $n+1$ rule.  

\begin{figure}[!htbp]
\begin{center}
\epsfig{file=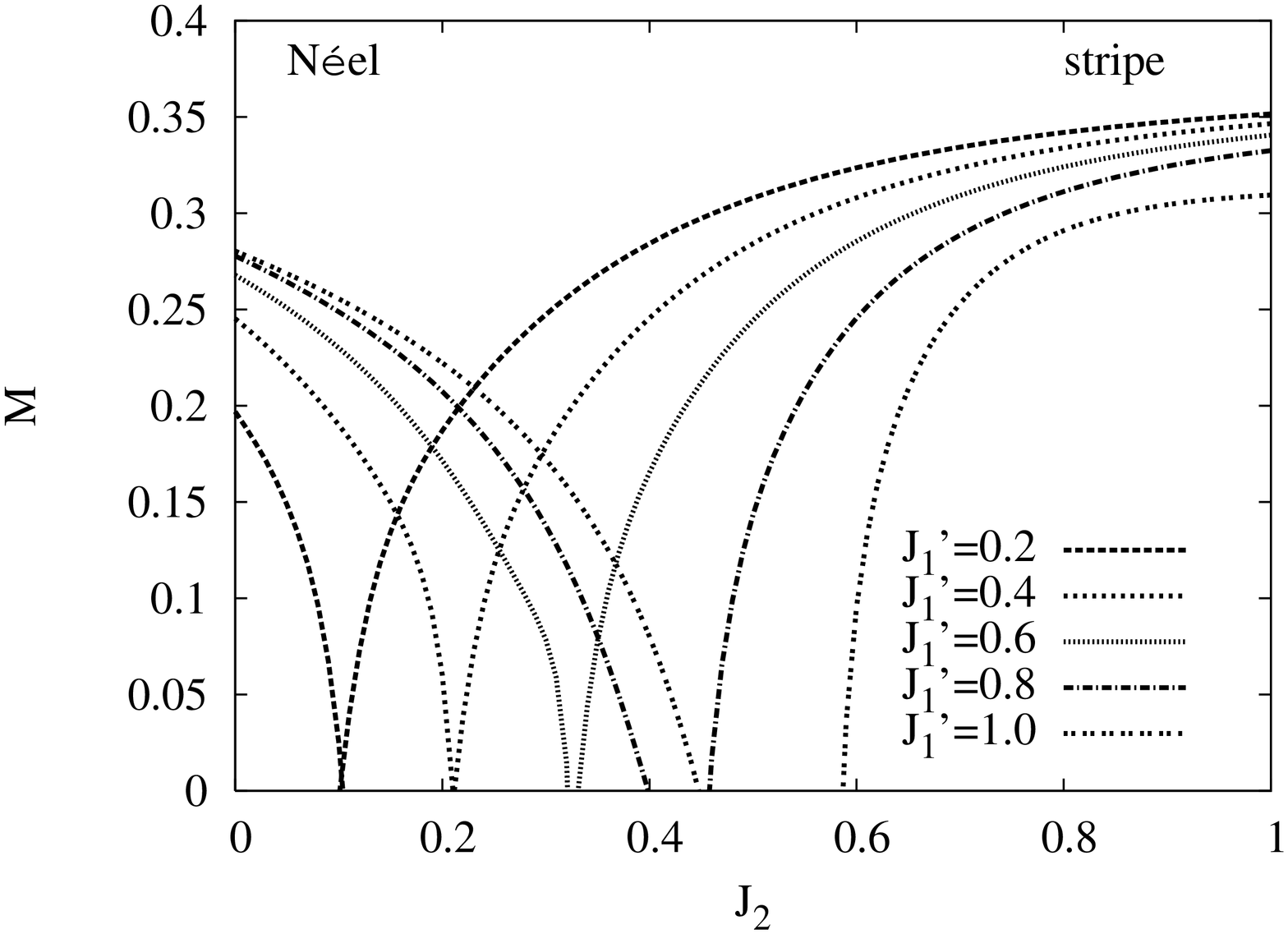,width=7cm,angle=270}
\caption{The extrapolated CCM LSUB$n$ results for the staggered magnetization, $M$, for $J_{1}'=0.2,0.4,0.6,0.8,1.0$ of the $s=1/2$ $J_{1}$--$J_{1}'$--$J_{2}$ model.  The LSUB$n$ results are extrapolated in the limit $n \rightarrow \infty$ using the set $n=\{4,6,8,10\}$.  The NN exchange coupling $J_{1}=1$.} 
\label{fig4}
\end{center}
\end{figure}

In figure~\ref{fig4}
we show our equivalent results for the staggered
magnetization to those in
figure~\ref{fig2} for the gs energy.  We note the surprising result that $M$ vanishes for both
the quantum N\'{e}el and stripe phases at almost exactly the same
critical value of $J_{2}$, for a given $J_{1}'$, so long as
$J_{1}'\lesssim0.6$.  Conversely, for $J_{1}'\gtrsim0.6$ there exists
an intermediate region between the critical points at which $M
\rightarrow 0$ for these two phases.  The order parameters of
both the N\'{e}el and the stripe phases vanish continuously both below
and above the correspondingly indicated quantum triple point (QTP), as is again typical of second-order
transitions.  We note, however, that there exists some evidence from
such other sources as ED calculations for the $J_{1}$--$J_{2}$ model (i.e., the present model with
$J_{1}'=J_{1}$) in 2D~\cite{Schulz:1996}, that the transition
between the stripe and intermediate phases is first-order, while
others~\cite{Su:2001} have argued it may be close to second-order.  We discuss these points further in Sec.~\ref{Discussion} below.  
     
We show in figure~\ref{fig5} 
\begin{figure}[t]
\begin{center}
\epsfig{file=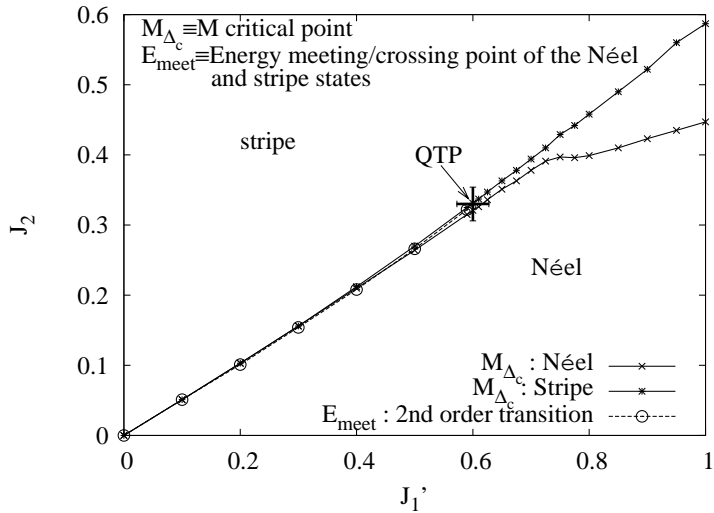,width=7cm,angle=270}
\caption{The extrapolated CCM LSUB$n$ results for the gs phase diagram of the $s=1/2$ $J_{1}$--$J_{1}'$--$J_{2}$ model.  The LSUB$n$ results are extrapolated in the limit $n \rightarrow \infty$ using the set $n=\{4,6,8,10\}$.  The NN exchange coupling $J_{1}=1$.  QTP $\equiv$ quantum triple point.}
\label{fig5}
\end{center}
\end{figure}
the zero-temperature phase diagram of the 2D spin-$1/2$
$J_{1}$--$J_{1}'$--$J_{2}$ model, as obtained from our extrapolated results for both the gs energy and the order parameter.  The phase diagram using the extrapolated LSUB$n$ results based on $n=\{2,4,6,8,10\}$ is very similar to that of figure~\ref{fig5}, which again adds credence to the validity and stability of our results.
  
Within the high-order CCM that we have used our results certainly seem to provide clear and consistent evidence for a QTP
at $J_{1}' \approx 0.6$ for ($J_{1}=1$).  For
$J_{1}'\lesssim0.6$ there exist only the N\'{e}el and stripe phases,
with a second-order transition between them.  For $J_{1}'\gtrsim0.6$
there exists an intermediate (magnetically disordered, i.e. paramagnetic) quantum phase, which requires further investigation.  Although the nature of
the intermediate phase is still under
discussion, a valence-bond solid phase seems to be the most
favoured~\cite{Ri:1993,Sir:2006}.  On the other hand, another possibility for the paramagnetic phase is the 
resonating valence bond (RVB) phase~\cite{Si:2004}. 

\section{Discussion and conclusions}
\label{Discussion}
In conclusion, our most important result is the evidence presented for a QTP
for which our best estimate is $J_{1}'\approx 0.60\pm0.03$, $J_{2}\approx 0.33\pm0.02$ for $J_{1}=1$.  Below this point we predict a second-order phase transition between the quantum N\'{e}el and stripe phases, whereas above it these two phases are separated by an intermediate phase.  Although the order parameters for the N\'{e}el and stripe phases vanish on the N\'{e}el-intermediate and stripe-intermediate phase boundary lines respectively above the QTP, we are unable to conclude more about the nature of the transitions at those boundaries, since the present calculations have not addressed at all the nature of the intermediate phase.  

Other calculations on this model~\cite{Si:2004,Star:2004} differ predominantly 
by giving a QTP at $J_{1}'=0=J_{2}$ for $J_{1}=1$.  We believe that the difference arises essentially from the nature of the alternative methods used.  For example, due to the small size of the lattices used, the ED calculations of Ref.~\cite{Si:2004} might miss the longer-range correlations that become increasingly important the nearer one approaches 
the QTP.  By contrast, at {\it any} level of truncation the CCM always incorporates some long-range correlations through the important exponentiated parametrization of the wave function that lies at the heart of the method, as described in Sec.~\ref{CCM}.  Also, the results of Starykh and Balents~\cite{Star:2004} are based on analysis of a continuum version of the $J_{1}$--$J_{1}'$--$J_{2}$ model rather than on the discrete lattice model itself.  This could easily account for our differing predictions for the position of the QTP.

As we have discussed above (and see figures~\ref{fig4} and \ref{fig5}), our results indicate that there is a continuous (second-order) transition below the QTP between the two ordered semiclassical phases with different broken symmetries.  This is undoubtedly a rather novel and surprising result that has few precedents, and one that seemingly violates the Landau criterion of symmetry change.  We note, however, that a similar scenario has been discussed recently in the related, but different, context of deconfined quantum criticality, as we explain more fully below.

Thus, it has been argued recently by other authors~\cite{Sen:2004} that there also exists a continuous (i.e., second-order) phase transition for the spin-1/2 pure $J_{1}$--$J_{2}$ model on the 2D square lattice (i.e., our case with $J_{1}'=J_{1}$) between the N\'{e}el state and what we have called here the intermediate state, and which those authors identify as a paramagnetic valence-bond solid (VBS) state.  Such direct second-order quantum phase transitions between two states with different broken symmetries, and hence characterized by two seemingly independent order parameters is difficult to understand within the standard Ginzburg-Landau critical theory, as we now discuss.

Thus, the competition between two such distinct kinds of quantum order associated with different broken symmetries would lead generically in the Ginzburg-Landau scenario to only one of three possibilities : (i) a first-order transition between the two states, (ii) an intermediate region of co-existence between both phases with both kinds of order present, or (iii) a region of intermediate phase with neither of the orders of these two phases present.  A direct second-order transition between states of different broken symmetries is only allowable within the standard Ginzburg-Landau critical theory if it arises by an accidental fine-tuning of the disparate order parameters to a multicritical point.  Thus, for the spin-1/2 pure $J_{1}$--$J_{2}$ model and its quantum phase transition suggested by Senthil {\it et al.}~\cite{Sen:2004}, it would require the completely accidental coincidence (or near coincidence) of the point where the magnetic order parameter (i.e., the staggered magnetization) vanishes for the N\'{e}el phase with the point where the dimer order parameter vanishes for the VBS phase.  Since each of these phases has a different broken symmetry (viz., spin-rotation symmetry for the N\'{e}el phase and the lattice symmetry for the VBS phase), one would naively expect both that each transition is described by its own {\it independent} order parameter (i.e.,the staggered magnetization for the N\'{e}el phase and the dimer order parameter for the VBS phase) and that the two transitions should therefore be {\it independent} of each other.

By contrast, the ``deconfined'' type of quantum phase transition postulated by Senthil {\it et al.}~\cite{Sen:2004} permits direct second-order quantum phase transitions between such states with different forms of broken symmetry.  In their scenario the quantum critical points still separate phases characterized by order parameters of the conventional (i.e., in their language, ``confining'') kind, but their proposed new critical theory involves fractional degrees of freedom (viz., spinons for the spin-1/2 $J_{1}$--$J_{2}$ model on the 2D square lattice) that interact via an emergent gauge field.  For the specific example of the spin-1/2 $J_{1}$--$J_{2}$ model the order parameters of both the N\'{e}el and VBS phases discussed above are represented in terms of the spinons, which themselves become ``deconfined'' exactly at the critical point.  That the spinons are the fundamental constituents of {\it both} order parameters then offers a natural explanation for the direct second-order phase transition between two states of the system that otherwise seem very different on the basis of their broken symmetries.

Despite the compelling nature of the arguments posited by Senthil {\it et al.}~\cite{Sen:2004}, we should mention, however, that other authors believe the phase transition in the $J_{1}$--$J_{2}$ model not to be due to a deconfinement of spinons.  For example, Sirker {\it et al.}~\cite{Sir:2006} have given arguments, based both on numerical results from series expansion analyses and on spin-wave theory, that the spin-1/2 $J_{1}$--$J_{2}$ transition is not of the above second-order deconfined type but, in their view, is more likely to be a (weakly) first-order transition between the N\'{e}el phase and a VBS phase with columnar dimerization.

One should also note that other, less radical, mechanisms have also been proposed to explain such second-order phase transitions and their seeming disagreement (except by accidental fine tuning) with Ginzburg-Landau theory.  What seems clearly to be minimally required is that the order parameters of the two phases with different broken symmetry should be related in some way.  Thus, a Ginzburg-Landau-type theory can only be preserved if it contains additional terms in the effective theory that represent interactions between the two order parameters.  For example, just such an effective theory has been proposed for the 2D spin-1/2 $J_{1}$--$J_{2}$ model on the square lattice by Sushkov {\it et al.}~\cite{Su:2002_PRB66}, and further discussed in Ref.~\cite{Sir:2006}.

Whether or not the deconfined phase transition theory of Senthil {\it et al.}~\cite{Sen:2004} survives the controversy that still surrounds it, we note that one of the motivations that led to it was the existence of various other numerical calculations in recent years that also point to a direct second-order quantum phase transition between phases of different broken symmetry that are characterized by seemingly independent order parameters.  Examples include the quantum phase transitions between : (i) an antiferromagnetic Mott insulator and a $d_{x^{2}-y^{2}}$ superconductor in a 2D Hubbard model on a square lattice, based on quantum Monte Carlo (QMC) simulations on lattices up to size $16\times16$~\cite{As:1997}; and (ii) superfluid and stripe-order phases in a 2D square-lattice spin-1/2 $XY$ model with both a nearest-neighbour coupling term ($J$) and a four-spin ring exchange term ($K$), based on a QMC method (viz., a stochastic series expansion technique) with lattices up to size $64 \times 64$~\cite{Sa:2002}.  While most previous such numerical evidence for continuous (second-order) quantum phase transitions between states with different broken symmetry has come from QMC simulations, the well-known ``sign problem'' inherent to QMC techniques has meant that it has not been easy to apply that method to frustrated spin-lattice systems of the type considered here.  We believe that the use of the CCM for such systems, as reported here, opens up a new arena and sheds a new spotlight on this fascinating and still unresolved larger field.

Returning to our own results for the spin-1/2 $J_{1}$--$J_{1}'$--$J_{2}$ model on the 2D square lattice, of course, one may also argue that what we have observed as a continuous (second-order) transition below the QTP may in reality be a very weak first-order transition, which is thereby not in violation of the Landau symmetry change criterion.  Our completely independent calculations for the two semiclassical phases can, obviously, never entirely preclude this latter possibility.  Nevertheless, it is clear from our results (and see figures~\ref{fig2} and \ref{fig4}) that the data below the QTP are fully consistent only with a transition which if it is not continuous is {\it very} weakly first-order for {\it all} values $0 \leq J_{1}' \lesssim 0.60$ below the QTP.

Finally, one could also argue that in this same range there might exist a very narrow strip of intermediate quantum-disordered phase, which would then reconcile our results (at least qualitatively) with those from the exact diagonalization of small clusters by Sindzingre~\cite{Si:2004}.  While this possibility, again, can never be entirely ruled out by any numerical calculation such as ours, we have clearly demonstrated that our own extrapolation schemes are both robust and internally consistent enough to rule out any but an {\it extremely} narrow strip of intermediate disordered phase for $0 \leq J_{1}' \lesssim 0.60$.  

We end by noting that two of the unique strengths of the CCM are its ability to deal with highly frustrated systems as easily as unfrustrated ones, and its use from the outset of infinite lattices, which leads in turn to its ability to yield accurate phase boundaries even near a possible
QTP.  Our own results for the ground-state energy and staggered magnetization provide a set of independent checks that lead us to believe that we have a self-consistent and coherent description of this extremely challenging system.  We also believe that the present high-order CCM results are among the numerically most accurate for this and related spin-lattice models containing frustration.  Nevertheless, our suggested novel phase scenario certainly needs further confirmation by the application of alternative high-order methods.  It would also be of considerable interest to repeat the investigation for the computationally more challenging case of the same system for spin-1 particles, and we intend to report on this system in the future.

\end{document}